\documentstyle[seceq,epsf,twoside]{ptptex}
\notypesetlogo  
\title{Covariant Classification Scheme of Hadrons}
\author{%
Shin {\sc Ishida} and Muneyuki {\sc Ishida}$^{*}$  }
\inst{%
Atomic Energy Research Institute, 
College of Science and Technology\\
Nihon University, Tokyo 101-0062, Japan\\
$^*$ Department of Physics, Tokyo Institute of Technology\\
Tokyo 152-8551, Japan
}
\recdate{%
\today
}
\abst{%
Starting from the multi-local Klein-Gordon equations
with Lorentz-scalar squared-mass operator we give a covariant
quark representation of the general composite mesons and baryons
with definite Lorentz transformation property. 
The mass spectra satisfy the approximate symmetry under the $\tilde U(4)$ 
transformation group, including the chiral transformation as a subgroup, 
concerning the spinor freedom of light constituent quarks, and this symmetry 
predicts the existence of new type of chiral mesons and baryons out of the 
conventional framework in non-relativistic quark model:
For example,
for light $q\bar q$ systems, the scalar $\sigma$- and axial-vector 
$a_1$-nonets, and for heavy-light $Q\bar q$ and $q\bar Q$ systems 
the scalar and axial-vector mesons are predicted to exist as relativistic
$S$-wave states besides the ordinary $P$-wave state mesons.
The existence of two ``exotic'' $1^{-+}$ meson nonets is predicted as the 
relativistic $P$-wave states in $q\bar q$ systems.  
For light quark baryons 
the extra \underline{\bf 56} with positive parity and the 
extra \underline{\bf 70} with negative parity of the static 
$SU(6)$ are predicted to exist as the ground state chiral particles.
}

\begin{document}
\maketitle

\setcounter{tocdepth}{4}

\section{Introduction}

There exist the two contrasting, non-relativistic and relativistic, viewpoints
of level-classification. The former is based on the non-relativistic quark model (NRQM)
with the approximate $LS$-symmetry and gives a theoretical base to the PDG 
level-classification. 
The latter is embodied typically in the NJL model with the approximate chiral symmetry.
It is widely accepted that $\pi$ meson nonet has the property as a Nambu-Goldstone
boson in the case of spontaneous breaking of chiral symmetry.

\begin{table}
\begin{center}
\begin{tabular}{l|c|c}
\hline
 & {\bf Non-Relativistic} & {\bf Relativistic} \\
\hline
 Model & Non Relat. Q. M. & NJL model \\
\hline
 Approx. Symm. & $LS$-Symm. & Chiral Symm. \\
\hline
 Evidence & Bases for PDG & $\pi$ nonet as NG boson\ \ \ \ \ \ \ \   \\
\hline
\end{tabular}
\end{center}
\caption{Two Contrasting Viewpoints of Level Classification}
\label{tab1}
\end{table}

Owing to the recent progress, both theoretical and experimental,
the existence of light $\sigma$-meson as chiral partner of $\pi (140)$
seems to be established\cite{rf1} especially through the analysis of various
$\pi\pi$-production processes. This gives further a strong support to
the relativistic viewpoint.

Thus, the hadron spectroscopy is now confronting with a serious problem,
existence of the seemingly contradictory two viewpoints, 
Non-relativistic and Extremely Relativistic ones.
The purpose of this talk is to present an attempt for a new 
level-classification scheme unifying these two viewpoints. 
The following is an overview of our attempt,
taking an example of the light-quark hadron system:

\begin{table}
\begin{center}
\begin{tabular}{l|c|@{}c@{}}
\hline
  & {\bf NRQM} & {\bf COQM} \\
  & & 
$\begin{array}{@{}cc} \ \ \ \ \ \ \ \ {\bf (old)} 
  & \ \ \ \ \ \ \ \ \ \ \ \ \ {\bf (extended)} \end{array}$ \\ 
\hline
 Confining force & spin-indep. 
     & 
 $\begin{array}{c|c} {\rm ``boosted"\ spin\ indep.}\ \ \ \ 
         & {\rm Lorentz\ scalar}  
  \end{array}$   \\
\hline
 Symmetry & $SU(6)_{SF}$ 
    & 
 $\begin{array}{c|c} {\rm ``boosted"}SU(6)_{SF}
         & \tilde U(12)_{SF} 
      \end{array}$ 
  \\
\hline
({\bf Wave Function}) & & $f(x_1,x_2)\ u_q\bar v_{\bar q}$\ \ 
{\it Dirac\ spinor}  \\
\ \ \ $(q\bar q)$-meson & 
    $f(${\bf x}$_1$,{\bf x}$_2)\ \chi_q\bar\chi_{\bar q}$\ \ 
{\it Pauli\ spinor}
 &
 $\begin{array}{c|c} u_q=u_+({\bf P}), & u_-(-{\bf P}) \\
         \bar v_{\bar q}=\bar v_+({\bf P}), & \bar v_-(-{\bf P}) \\
        boosted\ Pauli\ spin &   \end{array}$  
  \\
\ \ \ $(qqq)$-baryon & $f(${\bf x}$_1$,{\bf x}$_2$,{\bf x}$_3$)\ 
      $\chi_{q_1}\chi_{q_2}\chi_{q_3}$
 & $f(x_1,x_2,x_3)\ u_{q_1}u_{q_2}u_{q_3}$   \\
 & &  
 $\begin{array}{c|c} \ \ \ \ \ \ \ \ \ u_q=u_+({\bf P}), & u_-(-{\bf P}) \end{array}$
  \\
\ \ \ $(\bar q\bar q\bar q)$-anti-baryon
  & $f(${\bf x}$_1$,{\bf x}$_2$,{\bf x}$_3$)\ 
    $\bar \chi_{\bar q_1}\bar \chi_{\bar q_2}\bar \chi_{\bar q_3}$
  & $f(x_1,x_2,x_3)\ \bar v_{\bar q_1}\bar v_{\bar q_2}\bar v_{\bar q_3}$   \\
  & & 
 $\begin{array}{c|c} \ \ \ \ \ \ \ \ \ \bar v_{\bar q}=\bar v_+({\bf P}), & 
       \bar v_-(-{\bf P})
 \end{array}$
  \\
\ \ \ {\bf space(-time)} & 
  $\underline{O(3)\bigotimes SU(2)_S}\bigotimes SU(3)_F$
& $\underline{O(3,1)\bigotimes \tilde U(4)_{D.S.}}\bigotimes SU(3)_F$ \\
 & & \\
{\bf (Spin Wave Function)} & {\bf multi-Pauli spinor}
  & 
 $\begin{array}{c|c}  {\bf multi-}boosted\ \  & \ \ \ \ \ \ \ \  \\  
                      Pauli\ {\bf spinor}\ \  & \ \ \ \ \ \ \ \   \\ 
 \end{array}$ 
 \\
 & & \ \ \ \ \ \ \ \ \ \ \ \ \ \ \ {\bf multi-Dirac\ spinor} \\
\hline
\end{tabular}
\end{center}
\caption{Overview of Attempt : example of light quark hadrons}
\label{tab1}
\end{table}

The left column concerns the NRQM, while the right column does the 
covariant oscillator quark model (COQM) as a basic kinematical framework
of our attempt.

In NRQM the confining force is assumed to be spin-independent and the
mass spectra have the $SU(6)_{SF}$ spin-flavor symmetry.
In the extended (old) version of COQM the confining force is assumed to
be Lorentz-scalar (``boosted-spin" independent) and the mass spectra have the 
$\tilde U(12)_{SF}$ symmetry\footnote{
The $\tilde U(12)_{SF}$ symmetery was first proposed in 1965 as a
generalization of the static $SU(6)_{SF}$ symmetry. However, at that time
only the boosted Pauli-spinors are taken as physical components of fundamental 
representation of $\tilde U(4)_{D.S.}$. Now in the extended scheme
all general Dirac spinors
prove to be physical. } 
(boosted $SU(6)_{SF}$ symmetry).
The 3-dimensional space-coordinates of constituent quarks and$/$or anti-quarks,
as variables in the meson and baryon wave functions (WF), in NRQM are 
extended to the 4-dimensional Lorentz-vectors in COQM.
Similarly the multi-Pauli-spinors, as spin WF, in NRQM are extended to the
covariant multi-Dirac spinors in COQM;
where in the old version only the positive-energy spinors 
$u_+$({\bf P})($\bar v_+$({\bf P})) for quarks (anti-quarks) are considered, 
and in the extended version the negative-energy spinors 
$u_-$(-{\bf P})($\bar v_-$(-{\bf P})) 
(the {\bf P} is the space-momentum of hadrons as a whole-entity) are also 
taken into account. 
Thus, the WF of hadrons in the new level-classification scheme become the tensors
in the $O(3,1)_{\rm Lorentz}\bigotimes \tilde U(4)_{D.S.}\bigotimes SU(3)_F$-space
(, being extended from the ones in the $O(3)\bigotimes SU(2)_{P.S.}\bigotimes SU(3)_F$
space of NRQM).
The numbers of freedom of spin-flavor WF in NRQM 
are ${\bf 6}\times {\bf 6}^*=\underline{\bf 36}$
for mesons and $({\bf 6}\times {\bf 6}\times {\bf 6})_{\rm Symm.}
=\underline{\bf 56}$ for baryons:
These numbers in COQM become 
${\bf 12}\times {\bf 12}^*=\underline{\bf 144}$ for mesons and 
 $({\bf 12}\times {\bf 12}\times {\bf 12})_{\rm Symm.}=\underline{\bf 364}
=\underline{\bf 182}$ (for baryons)
$+\underline{\bf 182}$ (for anti-baryons).

Inclusion of heavy quarks is straightforward: The WF of general $q$ and$/$or $Q$ hadrons
become tensors in $O(3,1)\bigotimes [\tilde U(4)_{D.S.}\bigotimes SU(3)_F]_q
\bigotimes [SU(2)_{P.S.}\bigotimes U(1)_F]_Q$.

\section{Covariant framework for describing composite hadrons}

As WF of mesons and baryons we set up the following field-theoretical
expressions, respectively, as 
\begin{eqnarray}
\Phi_A{}^B(x_1,x_2) &=& \langle 0 | \psi_A(x_1) \bar\psi^B(x_2) | M \rangle 
              + \langle \bar M | \psi_A(x_1) \bar\psi^B(x_2) | 0 \rangle ,\\
\Phi_{A_1A_2A_3}(x_1,x_2,x_3) &=&  
\langle 0 | \psi_{A_1}(x_1) \psi_{A_2}(x_2)\psi_{A_3}(x_3)  | B
       \rangle   \nonumber\\ 
 & & + \langle \bar B | \psi_{A_1}(x_1) \psi_{A_2}(x_2)\psi_{A_3}(x_3)  | 0 
       \rangle,   
\label{eq2}
\end{eqnarray}
where $\psi_A$ is the quark field ($A=(\alpha ,a);\ \alpha = 1\sim 4\ \  (a)$ denoting
Dirac spinor (flavor) indices) and $\bar \psi^B$ denotes its Pauli-cpnjugate.
We start from the Yukawa-type Klein Gordon equation 
as a basic wave equation\cite{rf2}.
\begin{eqnarray}
[  \partial^2/\partial X_\mu^2   &-& 
 {\cal M}^2 ( r_\mu ,\partial / \partial r_\mu  ) ] \Phi (X,r,\cdots ) =0 ,               
\end{eqnarray}
where $X_\mu (r_\mu )$ are the center of mass (relative) coordinates of hadron systems.
The WF are separated into the positive (negative)-frequency parts concerning the CM 
plane-wave motion and expanded in terms of eigen-states of the squared-mass operator as
\begin{eqnarray}
\Phi (X,r,\cdots) &=& \sum_{P_N,N} 
\left[  e^{iP_N\cdot X} \psi_N^{(+)} (P_N,r,\cdots )
     +e^{-iP_N\cdot X} \psi_N^{(-)} (P_N,r,\cdots ) \right]   , \\
  & & {\cal M}^2 ( r_\mu ,\partial /i \partial r_\mu , \cdots )  \psi_N^{(\pm )} 
      =     M_N^2    \psi_N^{(\pm )}  , \\
  & & {\cal M}^2 = {\cal M}^2_{\rm conf} +\delta {\cal M}^2_{\rm pert.\ QCD} .
\end{eqnarray}
The ${\cal M}^2$ consists of two parts:
The confining-force part ${\cal M}^2_{\rm conf.}$ is assumed to be Lorentz-scalar
and $A,(B)$-independent, leading to the mass spectra with the $\tilde U(12)$
symmetry and also with the chiral symmetry. As its concrete model we apply the 
covariant oscillator in COQM, leading to the straight-rising Regge trajectories.
The effects due to perturbative QCD $\delta {\cal M}^2$ are neglected in this talk.

The internal WF is, concering the spinor freedom, expanded in terms of complete
set of relevant multi-spinors, Bargmann-Wigner (BW) spinors.
\begin{eqnarray}
{\rm meson}: &\  & \psi^{(\pm )}_{N,A}{}^B (P_N,r) = \sum_W 
W_\alpha^{(\pm )\beta }(P_N) M^{(\pm ) b}_{N,a} (r,P_N)\\
{\rm baryon}: &\  & \psi^{(\pm )}_{N,A_1A_2A_3} (P_N,r_1,r_2) = \sum_W 
W_{\alpha_1\alpha_2\alpha_3}^{(\pm )}(P_N) B_{N,a_1a_2a_3}(r_1,r_2,P_N) .\ \ \ 
\end{eqnarray}
The BW spinors are defined as multi-Dirac spinor solutions of 
the relevant local Klein-Gordon equation:
\begin{eqnarray}
(  \partial^2/\partial X_\mu^2 &-& M^2  ) 
  W_{\alpha\cdots}^{\ \ \beta\cdots} (X) = 0\\  
W_{\alpha\cdots}^{\ \ \beta\cdots} (X) & \equiv & \sum_{{\rm {\bf P}},P_0=E}
(e^{iPX}W_{\alpha\cdots}^{(+)\beta\cdots}(P)+ e^{-iPX}W_{\alpha\cdots}^{(-)\beta\cdots}(P)  ) .               
\end{eqnarray}
For mesons and baryons BW spinors are bi-Dirac and tri-Dirac spinors, respectively.
We further go into more details of BW spinors. First we define the Dirac spinors for
constituent quarks and anti-quarks with hadron 4-momentum $P_\mu$ as ``mono-index"
BW spinors:
\begin{eqnarray}
(  \partial^2/\partial X_\mu^2 - M^2  )  \psi_{\alpha} (X) &=& 0 , \\  
\psi_{q,\alpha} (X)  &\equiv&  \sum_{ P_\mu \  (P_0=\pm E_{\rm {\bf P}}) }
   e^{iPX} u_{q,\alpha}(P_\mu ) \nonumber\\
       &=& \sum_{{\rm {\bf P}},P_0=E_{\rm {\bf P}}}
(  u_+({\rm {\bf P}}) e^{iPX}  + u_-({\rm -{\bf P}}) e^{-iPX} ) , \\
\psi_{\bar q,\alpha} (X)  &\equiv&  \sum_{ P_\mu \   (P_0=\pm E_{\rm {\bf P}} ) }
   e^{-iPX} v_{\bar q,\alpha}(P_\mu ) \nonumber\\
        &=&  \sum_{{\rm {\bf P}},P_0=E_{\rm {\bf P}}}
(  v_+({\rm {\bf P}}) e^{-iPX}  + v_-({\rm -{\bf P}}) e^{iPX} ) , 
\end{eqnarray}
where the hadron 4 momentum $P_\mu$ satisfies the equations 
\begin{eqnarray}
P_\mu^2+M^2 &=& 0,\  P_0=\pm E_{\rm {\bf P}}, \  
E_{\rm {\bf P}}\equiv \sqrt{ {\rm {\bf P}}^2+M^2} .
\end{eqnarray}
Here it is to be noted that all 4-independent 
solution $u_q(P)$  ($v_{\bar q}(P)$) with
spin $\sigma_3(\sigma_3^\prime =-\sigma_3^T)=\pm 1$ 
and $P_0=\pm E_{\rm {\bf P}}$  
for quarks(anti-quarks) inside of hadrons.

The BW equations, the BW-spinors as their solutions and their irreducible composite
hadrons are summarized in Tables III and IV, respectively, for $q\bar q$-mesons and
$qqq$-baryons.
It is worthwhile to note that here exist new types of BW spinors for mesons(baryons);
$C(P),\ D(P)$ and $V(P)$ ($V(P)$ and $F(P)$) in addition to the conventional $U(P)\ (E(P))$,
boosted multi-Pauli spinors.

\begin{table}
\begin{center}
\begin{tabular}{lccc}
\hline
[Meson]\ \ \  $W_\alpha^{(+)\beta}(P)$ & $M^{(+)}(P)$ & BW-Equation
                   & ($P_0\equiv E_{\rm {\bf P}}>0)$ \\
\hline
 $U_\alpha{}^\beta (P)\equiv u_\alpha(P)\bar v^\beta (P);$ & $P_s,\ V_\mu .$ 
   & $(iP\cdot\gamma^{(1)}+M)U=0,$ & $U(-iP\cdot\gamma^{(2)}+M)=0$ \\
 $C_\alpha{}^\beta (P)\equiv u_\alpha(P)\bar v^\beta (-P);$ & $S,\ A_\mu .$ 
   & $(iP\cdot\gamma^{(1)}+M)C=0,$ & $C(iP\cdot\gamma^{(2)}+M)=0$ \\
 $D_\alpha{}^\beta (P)\equiv u_\alpha (-P)\bar v^\beta (P);$ & $S,\ A_\mu .$ 
   & $(-iP\cdot\gamma^{(1)}+M)D=0,$ & $D(-iP\cdot\gamma^{(2)}+M)=0$ \\
 $V_\alpha{}^\beta (P)\equiv u_\alpha (-P)\bar v^\beta (-P);$ & $P_s,\ V_\mu .$ 
   & $(-iP\cdot\gamma^{(1)}+M)V=0,$ & $V(iP\cdot\gamma^{(2)}+M)=0$ \\
\hline
\end{tabular}
\end{center}
\caption{Bargmann-Wigner (BW) Equations and Spinors for mesons.
Only the positive frequency parts 
are given.
The negative frequency parts $W^{(-)}$ are 
obtained from 
the operation $W^{(-)}=W^{(+)}\{ u\leftrightarrow v \}$.
For example, $C^{(-)}=C^{(+)} \{ u\leftrightarrow v \}=v_\alpha (P) \bar u^\beta (-P)$.
  }
\label{tab3}
\end{table}

\begin{table}
\begin{center}
\begin{tabular}{lcc}
\hline
[Baryon]\ \ \  $W^{(+)}_{\alpha_1\alpha_2\alpha_3}(P)$ & $B^{(+)}(P)$ & BW-Equation
                   ($P_0\equiv E_{\rm {\bf P}}>0)$ \\
\hline 
  $E_{\alpha_1\alpha_2\alpha_3}(P)\equiv u_{\alpha_1}(P)u_{\alpha_2}(P)u_{\alpha_3}(P);$
   & $\psi(\frac{1}{2}),\ \psi_\mu (\frac{3}{2}) .$ 
   & $(iP\cdot\gamma^{(1,2,3)}+M)E=0,$  \\
     & & \\
  $G_{\alpha_1\alpha_2\alpha_3}(P)\equiv u_{\alpha_1}(P)u_{\alpha_2}(P)u_{\alpha_3}(-P);$
   & $\psi(\frac{1}{2}),\ \psi_\mu (\frac{3}{2}) .$ 
   & $(iP\cdot\gamma^{(1,2)}+M)G=0,$ \\ 
     & & $(-iP\cdot\gamma^{(3)}+M)G=0,$   \\
  $F_{\alpha_1\alpha_2\alpha_3}(P)\equiv u_{\alpha_1}(P)u_{\alpha_2}(-P)u_{\alpha_3}(-P);$
   & $\psi(\frac{1}{2}),\ \psi_\mu (\frac{3}{2}) .$ & $(iP\cdot\gamma^{(1)}+M)G=0,$ \\
     & & $(-iP\cdot\gamma^{(2,3)}+M)G=0,$   \\
\hline
\end{tabular}
\end{center}
\caption{Bargmann-Wigner (BW) Equations and Spinors for baryons.
Only the positive frequency parts 
are given.
The negative frequency parts $W^{(-)}$ are obtained 
from the operation $W^{(-)}=W^{(+)}\{ u\rightarrow v \}$.
For example, $E_{\alpha_1\alpha_2\alpha_3}^{(-)}=E_{\alpha_1\alpha_2\alpha_3}^{(+)}
 \{ u\rightarrow v \}=v_{\alpha_1} (P) v_{\alpha_2} (P) v_{\alpha_3} (P)$.
}
\label{tab4}
\end{table}

\section{Transformation rule for hadrons and Chiral symmetry}

By using the covariant quark representation of composite hadrons given above
we can derive automatically their rule for any (relativistic) symmetry transformation
from that of constituent quarks. The rules for chiral transformation of mesons and baryons
are, respectively, 
\begin{eqnarray}
{\rm meson}: & & \psi_A{}^B(P,r) \longrightarrow [e^{i\alpha^a\lambda^a/2\ \gamma_5}\psi (P,r)
e^{i\alpha^a\lambda^a/2\ \gamma_5}]_A{}^B , \\
{\rm baryon}: & & \psi_{A_1A_2A_3}(P,r_1,r_2) \longrightarrow [ \Pi_{i=1}^3 
e^{i\alpha^a\lambda^{a,(i)}/2\ \gamma_5^{(i)}}\psi (P,r_1,r_2) ]_{A_1A_2A_3} . 
\end{eqnarray}
The physical meaning of chiral transformation are clearly seen from the operations:
\begin{eqnarray}
u(P) & \stackrel{\gamma_5}{\longrightarrow} & u'(P)=\gamma_5 u(P) =u(-P);\ \ 
       u_{\pm }({\rm {\bf P}}) \stackrel{\gamma_5}{\longleftrightarrow} u_{\mp }(-{\rm {\bf P}}), \\
v(P) & \longrightarrow & v'(P)=\gamma_5 v(P) =v(-P);\ \ 
       \bar v_{\pm }({\rm {\bf P}}) \longleftrightarrow \bar v_{\mp }(-{\rm {\bf P}}) .
 \end{eqnarray}
That is, the chiral transformation transforms the members of relevant BW-spinors with
each other. Accordingly, if ${\cal M}^2$ operator is independent of Dirac indices,
the hadron mass spectra have effectively the $\tilde U(4)$ symmetry and also the chiral
symmetry.

For convenience of later discussions we note further on physical meaning of BW equations
and introduce the notion of ``exciton-quark''. 
That is, the BW spinors with the total hadron momentum $P_\mu$ and mass $M$
are equivalent to the product of free Dirac spinors of the exciton quark 
with momentum
$p_\mu^{(i)}\equiv \kappa^{(i)} P_\mu $ and mass $m^{(i)}\equiv 
\kappa^{(i)} M\ (\sum_i \kappa^{(i)} =1)$,
as is seen from the equations 
(in an example of the $U$-type ($E$-type) BW spinors
of meson(baryon) systems).
\begin{eqnarray}
{\rm meson}\ \ \ \ 
(iP\cdot\gamma^{(1)}+M)U(P)=0 & \stackrel{\times \kappa^{(1)}}{\longrightarrow} &
(ip^{(1)}\cdot\gamma^{(1)}+m^{(1)})U(P)=0 \nonumber\\
\ \ \ \ \ \ \ \ \ \ \ \ \ \ 
U(P)(-iP\cdot\gamma^{(2)}+M)=0 & \stackrel{\times \kappa^{(2)}}{\longrightarrow} &
U(P)(-ip^{(2)}\cdot\gamma^{(2)}+m^{(2)})=0\ \ \ \ \ \ \ \  \label{eq19}\\
\ \ \ \ \ \ \ \ \ \ \ \ \ \ 
p_\mu^{(1)}+p_\mu^{(2)}=P_\mu ; & & m^{(1)}+m^{(2)}=M, \\
{\rm baryon}\ \ \ \ 
(iP\cdot\gamma^{(i)}+M)E(P)=0 & \stackrel{\times \kappa^{(i)}}{\longrightarrow} &
(ip^{(i)}\cdot\gamma^{(i)}+m^{(i)})E(P)=0 \label{eq21}\\
\ \ \ \ \ \ \ \ \ \ \ \ \ \ 
p_\mu^{(1)}+p_\mu^{(2)}+p_\mu^{(3)}=P_\mu ; & & m^{(1)}+m^{(2)}+m^{(3)}=M .
\end{eqnarray}
The above consideration is valid through all ground-state and$/$or excited state hadrons:
Accordingly the mass $M_N$ of the $N$-th excited hadron 
with the 4-momentum $P_N$
is generally given as a sum of the $N$-th excited
mass $m_N$ of the exciton quark with the 4-momentum
$P_N^{(i)}\equiv \kappa^{(i)} P_N$.
\begin{eqnarray}
M_N &=& m^{(1)}_N+m^{(2)}_N+\cdots ,\ \ 
p_N^{(i)} = \kappa^{(i)}  P_N\ \ (\sum_i \kappa^{(i)} =1).
\end{eqnarray}

\section{Level structure of mesons}
\subsection{Phenomenological\ criterion\ for\ chiral\ symmetry}
Considering the physical meaning of BW equations (see, Eqs.~(\ref{eq19}) and (\ref{eq21}))
mentioned in the last section, we may set up the phenomenological criterion for chiral
symmetry being effective as 
\begin{eqnarray}
m_{q,N}^2 & \ll & \Lambda_{\rm conf.}^2 (\approx \Lambda_{\chi SB}^2) 
\approx 1 {\rm GeV}^2 . \label{eq24}
\end{eqnarray}
We can estimate the values of exciton light-quark mass $m_{q,N}$ 
by applying the following 
mass formulas for the light-light $n\bar n$-meson ($n=u\ {\rm or}\ d$) 
and the light-heavy
$n\bar Q$- and $Q\bar n$-meson systems($Q=c$ or $b$).
\begin{eqnarray}  
M_N^2 & = & M_0^2 + N \Omega ,\ \ M_N=m_{q,N} + m_{q(Q),N}\ \ 
                (m_{q,0}=m_q,\ m_{Q,0}=m_Q)\ \ \ \  \label{eq25}\\
M_N^2 & = & \left\langle \left(\sqrt{m_q^2+{\bf p}^2}
+\sqrt{m_{q(Q)}^2+{\bf p}^2}\right)^2
+V \right\rangle_N \nonumber\\
         &\equiv&  
\left(\sqrt{m_q^2+\Lambda_N^2}+\sqrt{m_{q(Q)}^2+\Lambda_N^2}\right)^2.
\label{eq26}
\end{eqnarray}
The equation (\ref{eq25}) is the conventional formula in COQM, where the $\Omega^{-1}$
is the inverse Regge-slope and 
the zero-th exciton quark mass $m_{q,0} (m_{Q,0})$ is identified with the
corresponding constituent-quark mass $m_q (m_Q)$. 
The equation (\ref{eq26}) comes from the standard
bound-state picture of hadrons,
where $V$ is the scalar confining potential and $\Lambda_N$
corresponds to the average
value of relative momentum $|{\bf p}|$ of constituent quarks 
in the $N$-th excited meson rest-frame.
The result of values of light-exciton quark masses, 
thus estimated using the values of 
$\Omega$ and constituent quark masses obtained in the preceding analyses,
is collected in Table V.

\begin{table}
\begin{center}
\begin{tabular}{lc|cccc}
\hline
   &  &  $n\bar n$ & $n\bar c$ & $n\bar b$ & \\
\hline
 $\Omega$ & $/$GeV & 1.1 & 2.0 & 4.6 & Chiral symm. \\
\hline
$m_{n,N}$ & $N=0$ & 0.38 & 0.38 & 0.38 & $\bigcirc$ \\
\   (GeV) & $N=1$ & 0.64 & 0.70 & 0.74 & $\triangle$ \\
                & $N=2$  & 0.83 & 0.95 & 1.07 & $\times$ \\
\hline
Chiral & symm. & $N\le 1$ & $N\le 0$ or 1 & $N\le 0$ or 1 &   \\
\hline
\end{tabular}
\end{center}
\caption{Light exciton-quark mass $m_{n,N}$ for mesons}
\label{tab5}
\end{table}

By inspecting the values of $m_{n,N}$ in Table V in relation to the criterion
Eq.~(\ref{eq24}) we are able to infer that the chiral symmetry concerning the
light quarks is valid (still effective) for the ground (first excited) state of 
$n\bar n$ and $n\bar Q$ meson systems, while the symmetry will prove invalid
from the $N$-th ($N\ge 2$) excited hadrons.

\subsection{Level structure of ground state mesons} 
In Table VI we have summarized the properties of ground state mesons
in the light and$/$or heavy quark systems.
It is remarkable that there appear new multiplets of the scalar and axial-vector
mesons in the $q$-$\bar Q$ and $Q$-$\bar q$ systems and that in the $q$-$\bar q$
systems the two sets (Normal and Extra) of pseudo-scalar and of vector 
meson nonets exist. 
It is also to be noted that the $\pi$ nonet ($\rho$ nonet) is assigned to the
$P_s^{(N)}\ (V_\mu^{(N)})$ state, whose spin WF is much changed from that
in NRQM. We call the new type of particles in the extended COQM
(which have never appeared in NRQM) as ``chiralons''.

\begin{table}
\begin{center}
\begin{tabular}{l|c|l|ccl}
\hline
   & mass & Approx. Symm. & Spin WF & $SU(3)$ & Meson Type \\
\hline
$Q\bar Q$ & $m_Q+m_{\bar Q}$ & $LS$ symm. & $u_Q(P)\bar v^{\bar Q}(P)$
          & $\underline{1}$ & $P_s,\ V_\mu$  \\
\hline
$q\bar Q$ & $m_q+m_{\bar Q}$ & $q$-Chiral Symm. & $u_q(P)\bar v^{\bar Q}(P)$
          & $\underline{3}$ & $P_s,\ V_\mu$  \\
               &                              & $\bar Q$-HQS  & $u_q(-P)\bar v^{\bar Q}(P)$
          & $\underline{3}$ & $S,\ A_\mu$  \\
$Q\bar q$ & $m_Q+m_{\bar q}$ & $\bar q$-Chiral Symm. & $u_Q(P)\bar v^{\bar q}(P)$
          & $\underline{3}^*$ & $P_s,\ V_\mu$  \\
               &                              & $Q$-HQS  & $u_q(P)\bar v^{\bar Q}(-P)$
          & $\underline{3}^*$ & $S,\ A_\mu$  \\
\hline
$q\bar q$ & $m_q+m_{\bar q}$ & Chiral Symm. & $(1/\sqrt{2})(u(P)\bar v(P)$
          & \hspace{-0.5cm}$\pm u(-P)\bar v(-P) )$ & $P_s^{(N,E)},\ V_\mu^{(N,E)}$  \\
              &                             &                     & $(1/\sqrt{2})(u(P)\bar v(-P)$
          & \hspace{-0.5cm}$\pm u(-P)\bar v(P) )$ & $S^{(N,E)},\ A_\mu^{(N,E)}$  \\
\hline
\end{tabular}
\end{center}
\caption{Level structure of ground-state mesons}
\label{tab6}
\end{table}

\subsection{Level structure of mesons in general} 
The mass of the ground and excited state mesons is given by
\begin{eqnarray} 
M_N^2 & = & M_0^2+N\Omega = m_N^{(1)} + m_N^{(2)} .
\end{eqnarray}
Their quantum numbers are given in Table VII.
Here it is to be noted that some chiralons have the ``exotic" quantum numbers
from the conventional NRQM viewpoint.

\begin{table}
\begin{center}
\begin{tabular}{lc|lc}
\hline
$(q\bar q)$  &  &  &  \\
\ \ $N=$all  &  $P_s^{(N,E)}\bigotimes \{ L,N \}$ & \ \ $P=(-1)^{L+1}$ & $C=(-1)^L$  \\
             &  $V_\mu^{(N,E)}\bigotimes \{ L,N \}$ & \ \ $P=(-1)^{L+1}$ & $C=(-1)^{L+1}$  \\
\ \ $N=0$(and 1)  &  $S^{(N,E)}\bigotimes \{ L,N \}$ & \ \ $P=(-1)^{L}$ & $C=\pm (-1)^L$  \\
             &  $A_\mu^{(N,E)}\bigotimes \{ L,N \}$ & \ \ $P=(-1)^{L}$ & $C=\pm (-1)^{L}$  \\
\hline
$(q\bar Q$ or $Q\bar q)$  &  &  $(Q\bar Q)$ &    \\
\ \ $N=$all  &  $P_s \bigotimes \{ L,N \}$ & \ \ $N=$all  &  $P_s \bigotimes \{ L,N \}$ \\
           &  $V_\mu \bigotimes \{ L,N \}$ &      &   $V_\mu \bigotimes \{ L,N \}$  \\
\ \ $N=0$(and 1)  &  $S \bigotimes \{ L,N \}$   &   &    \\
             &  $A_\mu \bigotimes \{ L,N \}$   &   &  \\
\hline
\end{tabular}
\end{center}
\caption{Level structure of Mesons in general}
\label{tab7}
\end{table}

The schematic picture of meson spectroscopy is shown in Fig. {\bf 1}.

\begin{figure}[t]
  \epsfxsize=12 cm
  \epsfysize=6 cm
 \centerline{\epsffile{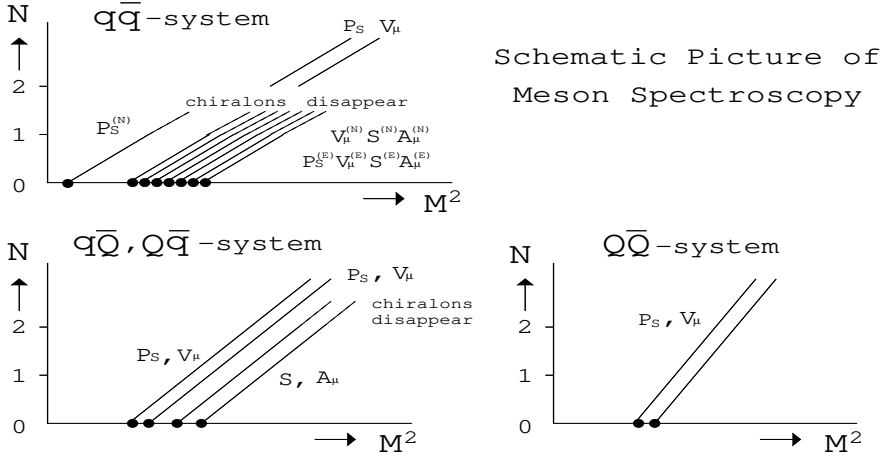}}
 \caption{Schematic picture of meson spectroscopy}
  \label{fig:1}
\end{figure}

\section{Level Structure of Baryons}

The baryon WF Eq.~(\ref{eq2}) should be full-symmetric (except for the color freedom)
under exchange of constituent quarks: The full-symmetric total WF in the extended
scheme is obtained, in the following three ways, as a product of the sub-space WF with
respective symmetric properties:
\begin{eqnarray}	
| \rho F \sigma \rangle_S & = & | \rho \rangle_S |F\sigma\rangle_S\ \ \ (a);\ \ \ \ 
 | \rho \rangle_\alpha |F\sigma\rangle_\alpha + | \rho \rangle_\beta |F\sigma\rangle_\beta\ \ \ (b);
\nonumber\\
 &  & | F \rangle_A |\rho\sigma\rangle_A\ \ \ (c);
\end{eqnarray}
where $|\rho\rangle_S$ is the full-symmetric $\rho$-spin space WF and so on.
$\rho\bigotimes\sigma =\gamma$ is the conventional two, $\rho$ and $\sigma$ spin,
2 by 2 matrix representation of the 4 by 4 Dirac matrix, and $|\ \  \rangle_{\alpha (\beta ),A}$
mean the $\alpha (\beta)$-type partial symmetric and full anti-symmetric subspace WF,
respectively.
The intrinsic parity operation is given by $\hat P = \Pi_{i=1}^3 \gamma_4^{(i)}$, that is,
the parity of $(E^{(+)},G^{(+)},F^{(+)})$ BW spinors are $(+,-,+)$ and those of
$(E^{(-)},G^{(-)},F^{(-)})$ BW spinors are $(-,+,-)$.
The symmetry properties of ground state light-quark baryon WF and their level
structures thus determined are summarized in Table VIII.  

\begin{table}
\begin{center}
\begin{tabular}{clcrc}
\hline
$W^{(+)}$ & spin-flavor wave function & $B^{P\hspace{-0.22cm}\bigcirc}$ 
  &  static & $SU(6)$ \\ 
\hline
$E^{(+)}$ : & 
$|\rho\rangle_S |F\sigma\rangle_S=|\rho\rangle_S |F\rangle_S|\sigma\rangle_S$ 
 & $\Delta_{3/2}^{+\hspace{-0.22cm}\bigcirc}$ & $10\times 4=40$ &   \\
  & \ \ \ \ \ \ \ \ \ \ \ \ \ \ \ \ \ 
$|\rho\rangle_S (|F\rangle_\alpha |\sigma\rangle_\alpha +
|F\rangle_\beta |\sigma\rangle_\beta )$ 
 & $N_{1/2}^{+\hspace{-0.22cm}\bigcirc}$ & $8\times 2=16$ & \underline{\bf 56} \\
\hline
$G^{(+)}$ : & 
$|\rho\rangle_\alpha |F\sigma\rangle_\alpha +|\rho\rangle_\beta |F\sigma\rangle_\beta$ ; 
$|F\sigma \rangle_{\alpha (\beta )} = |F\rangle_S |\sigma\rangle_{\alpha (\beta )}$
 & $\Delta_{1/2}^{-\hspace{-0.22cm}\bigcirc}$ & $10\times 2=20$ &   \\
  & \ \ \ \ \ \ \ \ \ \ \ \ \ \ 
\ \ \ \ \ \ \ \ \ \ \ \ \ \ \ \ \ \ \ \ \ \ \ \ \ \ \ \ \ \ \ \ \ \ 
$|F\rangle_{\alpha (\beta )} |\sigma\rangle_S$ 
 & $N_{3/2}^{-\hspace{-0.22cm}\bigcirc}$ & $8\times 4=32$ &    \\
 & 
$|F\rangle_A  |\rho\sigma\rangle_A=|F\rangle_A(-|\rho\rangle_\alpha |\sigma\rangle_\beta
   +|\rho\rangle_\beta |\sigma\rangle_\alpha )$  
 & $\Lambda_{1/2}^{-\hspace{-0.22cm}\bigcirc}$ & $1\times 2=2$ &   \\
  & 
$|\rho\rangle_S |F\sigma\rangle_S=
 |\rho\rangle_S (|F\rangle_\alpha |\sigma\rangle_\beta +
|F\rangle_\beta |\sigma\rangle_\alpha )$ 
 & $N_{1/2}^{-\hspace{-0.22cm}\bigcirc}$ & $8\times 2=16$ & \underline{\bf 70} \\
\hline
$F^{(+)}$ : & 
$|\rho\rangle_S |F\sigma\rangle_S=|\rho\rangle_S |F\rangle_S|\sigma\rangle_S$ 
 & $\Delta_{3/2}^{+\hspace{-0.22cm}\bigcirc}$ & $10\times 4=40$ &   \\
  & \ \ \ \ \ \ \ \ \ \ \ \ \ \ \ \ \ 
$|\rho\rangle_S (|F\rangle_\alpha |\sigma\rangle_\alpha +
|F\rangle_\beta |\sigma\rangle_\beta )$ 
 & $N_{1/2}^{+\hspace{-0.22cm}\bigcirc}$ & $8\times 2=16$ & \underline{\bf 56} \\
\hline
 & $_{12}H_3=\underline{\bf 364}\stackrel{\left(  \times\frac{1}{2}\right) }{\longrightarrow}
\underline{\bf 182}=\underline{\bf 56}^{+\hspace{-0.22cm}\bigcirc} 
+\hspace{-0.4cm}\bigcirc \underline{\bf 56}^{+\hspace{-0.22cm}\bigcirc} 
+\hspace{-0.4cm}\bigcirc \underline{\bf 70}^{-\hspace{-0.22cm}\bigcirc}$ & & & \\
 & \hspace{1cm}$\underline{\bf 56}^{+\hspace{-0.22cm}\bigcirc}$\ \ \ \ \ \ 
   $N_{1/2}^{+\hspace{-0.22cm}\bigcirc}$, 
  $\Delta_{3/2}^{+\hspace{-0.22cm}\bigcirc}$  & & & \\ 
 & $\begin{array}{|l|}\hline
   \hspace{0.8cm}\underline{\bf 56}^{+\hspace{-0.22cm}\bigcirc}\ \ \ \ \ \ 
   N_{1/2}^{\prime +\hspace{-0.22cm}\bigcirc}, 
  \Delta_{3/2}^{\prime +\hspace{-0.22cm}\bigcirc}
  \ \ \ \ \ \ \ \ \ \ \ \ \ {\bf chiralons}\ \ \ \ \ \\ 
       \hspace{0.8cm}\underline{\bf 70}^{-\hspace{-0.22cm}\bigcirc}\ \ \ \ \ \ 
       N_{1/2}^{-\hspace{-0.22cm}\bigcirc}, N_{3/2}^{-\hspace{-0.22cm}\bigcirc},
       \Delta_{1/2}^{-\hspace{-0.22cm}\bigcirc},
       \Lambda_{1/2}^{-\hspace{-0.22cm}\bigcirc}  \\ 
  \hline \end{array}$    & & & \\ 
\hline
\end{tabular}
\end{center}
\caption{Level structure of ground-state $qqq$-baryon}
\label{tab8}
\end{table}
 
Here it is remarkable that there appear chiralons in the ground states.
That is, the extra positive parity \underline{\bf 56}-multiplet of the static
$SU(6)$ and the extra negative parity \underline{\bf 70}-multiplet of the
$SU(6)$ in the low mass region. 
It is also to be noted that the chiralons in the first excited states are 
expected to exist.
The above consideration on the light-quark baryons are extended
directly to the general light and$/$or heavy quark baryon systems:
The chiralons are expected to exist also in the $qqQ$ and $qQQ$-baryons,
while no chiralons in the $QQQ$ system.

\section{Experimental Candidates for Chiral Particles}

In our level-classification scheme a series of new type of multiplets of the particles,
chiralons, are predicted to exist in the ground and the first excited states of $q\bar q$
and $q\bar Q$ or $Q\bar q$ meson systems and of $qqq$, $qqQ$ and $qQQ$-baryon
systems. Presently we can give only a few experimental candidates or indications for
them:\\
{\em ($q\bar q$-mesons)}\ \ \ \ One of the most important candidates is the scalar
$\sigma$ nonet to be assigned as $S^{(N)}(^1S_0)$ : $[\sigma (600),\ \kappa (900),\ 
a_0(980),\ f_0(980)]$. The existence of $\sigma (600)$ seems to be 
established\cite{rf1} through
the analyses of, especially, $\pi\pi$-production processes. 
A firm experimental evidence\cite{rf3} for $\kappa (800$--$900)$
through the decay process\cite{rf4} $D^+\rightarrow K^-\pi^+\pi^+$
was reported at this conference.\\
In our scheme respective two sets of $P_s$- and of $V_\mu$-nonets,
to be assigned as $P_s^{(N,E)}(^1S_0)$ and $V_\mu^{(N,E)}(^3S_1)$,
are to exist: Out of the five vector mesons (stressed\cite{rf5} 
as problems with vector mesons, 
$[\rho^\prime (1450),\  \rho^\prime (1700),\ 
\omega^\prime (1420),\  \omega^\prime (1600),\  \phi (1690) ]$,
the lower mass $\rho^\prime (1450)$ and $\omega^\prime (1420)$,
and the $\phi (1690)$ are naturally able to be assigned as the members
of $V_\mu^{(E)}$-nonets;\\
Out of the three established $\eta$, $[\eta (1295),\ \eta (1420),\  \eta (1460) ]$
at least one extra, plausibly $\eta (1295)$ with the lowest mass, may belong to
$P_s^{(E)}(^1S_0)$ nonet.\\
Recently the existence of two ``exotic" particles $\pi_1(1400)$ and $\pi_1(1600)$
with $J^{PC}=1^{-+}$ and $I=1$, observed\cite{rf6}  
in the $\pi\eta ,\  \rho\pi$ and other channels, is attracting strong interests
among us.
These exotic particles with a mass around 1.5GeV may be naturally assigned 
as the first excited states $S^{(E)}(^1P_1)$ and $A_\mu^{(E)}(^3P_1)$ of the
chiralons.\\
({\em $q\bar Q$}\ or\  $Q\bar q$-mesons)\ \ \ \ At this conference
some experimental indication for existence of the two chiralons in
$D$- and $B$-meson systems obtained through
the $\Upsilon (4S)$ or $Z^0$ decay process, were reported, respectively,\\
\hspace*{2cm}
$D_1^\chi = A_\mu (^3S_1), \ \  J^P=1^+\ \ {\rm in}\ \ D_1^\chi \rightarrow D^* + \pi$ \\
\hspace*{2cm}
$B_0^\chi = S (^1S_0), \ \  J^P=0^+\ \ {\rm in}\ \ B_0^\chi \rightarrow B + \pi  $,\\   
by Yamada K\cite{rf7} and Ishida M\cite{rf8}.\\
({\em $qqq$}-baryons)\ \ \ \  The two facts have been a longstanding problem
that the Roper resonance
$N(1440)_{1/2^+}$ is too light to be assigned
as radial excitation of $N(939)$ and that $\Lambda (1405)_{1/2^-}$ 
is too light as the $L=1$ excited state of $\Lambda (1116)$.
In our new scheme they are reasonably assigned to the members of ground
state chiralons with $[SU(6),\ SU(3),\ J^P]$, respectively, as \\
\hspace*{1cm}
$N(1440)_{1/2^+} = F(\underline{\bf 56},\ \underline{\bf 8},\  {1}/{2}^+),\ \ 
\Lambda (1405)_{1/2^-} = G(\underline{\bf 70},\ \underline{\bf 1},\   {1}/{2}^-)$\ .\\
The particle $\Delta (1600)_{3/2^+}$ which is lighter than 
$\Delta (1620)_{1/2^-}$ may also belong to the extra \underline{\bf 56}
of the ground state chiralons. This situation is shown in Table IX. 

\begin{table}
\begin{tabular}{c|cl|cl}
\hline
SU(6)  & SU(3), $J^P$ &  &  SU(3), $J^P$ &   \\
\hline
\underline{\bf 56} & \underline{\bf 8},\ $\frac{1}{2}^+$ 
     & $N(939),\Lambda (1116),\Sigma (1192),
\Xi (1318)$ & \underline{\bf 10},\ $\frac{3}{2}^+$
     & $\Delta (1232),\Sigma (1385),\cdots $ \\
\underline{\bf 56}$^\prime$ & \underline{\bf 8},\ $\frac{1}{2}^+$ & 
$\begin{array}{|c|}\hline N(1440)\\  \hline \end{array}$,\ \ \ \ \ \ \ 
$\begin{array}{|c|}\hline \Sigma (1660)\\  \hline \end{array}$ 
  & \underline{\bf 10},\ $\frac{3}{2}^+$ & 
$\begin{array}{|c|}\hline \Delta (1600)  \\  \hline \end{array}$   \\
\hline 
\underline{\bf 70} & \underline{\bf 8},\ $\frac{1}{2}^-$ & 
$N(1535)$  & \underline{\bf 10},\ $\frac{1}{2}^-$ & $\Delta (1620)$   \\
   & \underline{\bf 1},\ $\frac{1}{2}^-$ & 
---------$\begin{array}{|c|}\hline \Lambda (1405)\\  \hline \end{array}$-----------------  &  &   \\
\hline
\end{tabular}
\caption{Assignment of $qqq$-baryons: The baryons in the boxes are candidates of chiralons.}
\label{tab9}
\end{table}

\section{Concluding Remarks}

I have presented an attempt for Level-classification 
scheme \underline{unfying}
   the seemingly contradictory \underline{two\ viewpoints};
\underline{Non-relativistic\ one} with $LS$-symmetry and\\
\underline{Relativistic\ one} with Chiral symmetry . 

As results, I have predicted the existence of \underline{New Chiral Particles
    in the lower}\\
\underline{mass regions} ``Chiralons'', which had never been appeared 
in NRQM.

  We have several \underline{good candidates} for chiralons, for example,\\
$\sigma$-nonet\ \  \{  
                 $\sigma (600)$, $\kappa (800)$, $a_0(980)$, $f_0(980)$   \}
                {\em as ``Relativistic'' S-wave states of $(q\bar q)$}.\\ 
$\pi_1(1400)$,\  $\pi_1 (1600)$,\  $(1^{-+})$ ; 
                {\em as ``Relativistic'' P-wave states of $(q\bar q)$}. \\  
Roper resonance $N(1440)_{1/2^+}$ and 
             SU(3) singlet $\Lambda (1405)_{1/2^-}$;
     {\em as ``Relativistic'' S-wave states of $(qqq)$}.   

 Further search, both experimental and theoretical, for chiralons is necessary
and important.



\end{document}